\documentstyle[12pt,a4,epsf]{article}
\newfont{\elffont}{cmr10 scaled 1096}
\renewcommand{\baselinestretch}{1.5}
\date{}

\begin{document}
\vspace*{2.4cm}
\begin{center}
{\large ANOMALOUS ISOTOPE SHIFTS IN Pb NUCLEI IN RMF THEORY}\\
\       \\
\       \\
G.A. Lalazissis$^{1)}$, M.M. Sharma$^{2)}$, J. K\"onig$^{1)}$, P. Ring$^{1)}$\\
$^{1)}$Physikdepartment, Technische Universit\"at M\"unchen, D-85747 Garching\\
$^{2)}$Max-Planck Institut f\"ur Astrophysik, D-85740 Garching\\
\end{center}
\vspace{3cm}
{\renewcommand{\baselinestretch}{1.} \small\normalsize
\begin{abstract}

We have studied the anomalous behaviour of isotopic shifts
of Pb nuclei in the relativistic mean field theory.
It has been shown that the relativistic mean field
provides an excellent description of the anomalous kink in the isotopic
shifts about $^{208}$Pb. This is in contrast from density-dependent
Skyrme forces which do not reproduce the observed trend in
the empirical data on the charge radii. We discuss some differences in
the description of isotope shifts in the RMF theory and the
Skyrme mean field.

\end{abstract}}
\bigskip\smallskip
\centerline{ \bf 1. Introduction}

The relativistic mean field (RMF) theory$^{1,2)}$ has of late
been very successful in describing the ground-state properties of
nuclei at and away from the stability line. This is in contrast to
phenomenological density dependent Skyrme forces$^{3)}$ where
the description of nuclei is constrained mainly to the stability
line. Moreover, the spin-orbit interaction arising from
the Dirac description of nucleons provides an attractive feature
of the RMF theory. In the non-relativistic Skyrme theory
on the other hand, the spin-orbit interaction is added only
phenomenologically. The saturation mechanisms of the two theories are
evidently quite different. In the RMF theory the saturation and the
density dependence of the nuclear interaction stems from a balance
between large attractive scalar $\sigma$-meson field and large repulsive
vector $\omega$-meson field . The asymmetry component is accounted for by
the isovector $\rho$ meson. In essence, the nuclear interaction is
generated by the exchange of various mesons between nucleons. This leads
to the structure of the force in the RMF theory and consequently
a density dependence which differs from that of Skyrme theory.
In the Skyrme ansatz, however, the density dependent term
is phenomenological and is obtained by fitting the properties
of nuclei.

The charge radii of Pb isotopes and their isotope shifts have been
investigated in detail$^{4)}$ using density-dependent Skyrme
forces. The isotopic chain of Pb nuclei is known
to exhibit a kink in the empirical isotope shifts$^{5)}$ about
shell-closure. This implies that heavier isotopes obtained on adding
further neutrons to $^{208}$Pb show unusually large charge
radii as compared to the lighter ones. The Skyrme forces do succeed
in describing the isotopes shifts and thus charge radii of nuclei
only on the lighter side of $^{208}$Pb, where a density-dependent
pairing force is required to be switched on.
The isotope shifts of the heavier nuclei, can not, however, be
described by any of the standard Skyrme forces, as has been discussed
in detail in ref.$^{4)}$. Even including possible ground-state
correlations does not improve the description.
Thus, the Skyrme mean field is not able to describe the charge radii of
heavier isotopes of Pb. In this work, we investigate the long-standing
problem of the isotope shifts in the RMF theory.
In sect. 2 we present briefly some features of the RMF theory.
In sect. 3 our results are presented and discussed.\\

\centerline { \bf 2.  The RMF theory approach}

The ansatz of the interaction in the RMF theory is based upon
Lagrangian density of the form$^{1)}$:

\begin{equation}
\begin{array}{rl}
{\cal L} =&
\bar \psi (i\rlap{/}\partial -M) \psi +
\,{1\over2}\partial_\mu\sigma\partial^\mu\sigma - U(\sigma)
-{1\over 4} \Omega_{\mu\nu}\Omega^{\mu\nu} +
   {1\over2}m_\omega^2\omega_\mu\omega^\mu
  -{1\over4}{\vec R}_{\mu\nu}{\vec R}^{\mu\nu}\\
\ &   + {1\over2}m_\rho^2\vec\rho_\mu\vec\rho^\mu
     -{1\over 4}F_{\mu\nu}F^{\mu\nu}
      -g_\sigma \bar\psi \sigma \psi~
    -~g_\omega \bar\psi \rlap{/}\omega \psi~
    -~g_\rho   \bar\psi \rlap{/}\vec\rho\vec\tau\psi
    -~e        \bar\psi A\llap{/}\psi.
\end{array}
\end{equation}
\medskip
\par\noindent
where the Dirac nucleon interacts with the $\sigma$ and $\omega$ meson
fields. The $\rho$ meson generates the isovector component of the force.
The nonlinear $\sigma\omega\rho$ model which we use, has a nonlinear
scalar self-interaction of the $\sigma$ mesons as given by

\begin{equation}
U(\sigma)~=~{1\over 2}m_\sigma^2\sigma^2~+~
{1\over 3}g_2\sigma^3~+~{1\over 4}g_3\sigma^4,
\end{equation}
where $g_2$ and $g_3$ are the non-linear parameters. Details on the
RMF theory have been discussed in ref.$^{2)}$. The parameter sets NL1$^{6)}$
and NL2$^{7)}$ have been used extensively to obtain properties of nuclei.
It was shown by Sharma and Ring$^{8)}$ that both the above forces provide
neutron skin thickness of neutron-rich nuclei much larger than
the empirical values. Investigating the ground-state properties of
nuclei$^{9)}$ in the non-linear $\sigma\omega\rho$ model, it was noted
that indeed a stronger $\rho$ meson coupling and therefore a very large
asymmetry energy of the above forces has been responsible for
larger neutron skin thickness of neutron-rich nuclei. Consequently,
a new force NL-SH was obtained, where the above problem of
the earlier forces was resolved. It was also shown$^{9)}$ that this
force describes very well the ground-state binding energies, charge and
neutron radii of spherical nuclei near the stability line
as well as those of deformed nuclei very far off the stability line.
Here, we study isotope shifts in the RMF theory using the forces NL1 and
NL-SH.\\

\centerline {\bf 3. Results and discussion}

We have performed calculations within the Hartree approximation.
Although most of the Pb isotopes close to $^{208}$Pb are spherical,
an axially symmetric configuration has been assumed and Hartree
minimization has been performed. The method of the oscillator expansion$^{10)}$
has been employed, whereby both the fermionic as well as bosonic
wavefunctions have been expanded in N = 12 shells. We have considered
all the even-mass Pb isotopes from A = 190 to 214. For convergence reasons,
N = 14 Fermionic shells have also been considered. It is found
that the difference between the N=12 and N=14 calculations is very small.
Therefore, only the results obtained with N=12 are presented.
For all the open-shell nuclei pairing has been included within the BCS
formalism. The pairing gaps have been obtained from the particle separation
energies of the neighbouring nuclei. The quadrupole deformations obtained
from the convergence are very small.  These are practically close to spherical
configuration for all the nuclei we have considered.

The binding energy per nucleon of Pb isotopes obtained in
the RMF theory are shown in Fig 1a. The empirical binding energies are
also shown for comparison. The binding energies of all isotopes are
reproduced well by the set NL-SH. The deviations are
at most 0.1\%. Here we also compare the results obtained with NL1.
For the lighter nuclei of the isotopic chain, the binding
energies obtained with NL1 show a systematic deviation from
the empirical data. The difference between the calculated
and the empirical binding energies shows an increase as the neutron
excess decreases. This discrepancy is due to the asymmetry
energy of about 44 MeV for NL1, which is larger than the empirical value.


The calculated charge radii have been used to obtain the isotope
shifts. The nucleus $^{208}$Pb has been taken as the reference point.
In order to provide a good illustration,
the isotope shifts ($\Delta r_c^2 = r_c^2(A) - r_c^2(208)$)
have been modified by substracting an equivalent of the liquid-drop
difference ($ \Delta r_{LD}^2 = r_{LD}^2(A) - r_{LD}^2(208)$) obtained
from $r^2_{LD}(A) = {3\over 5}r_0^2A^{2/3}$ for Pb nuclei, as in ref.$^{4)}$.
All the results are presented in the same way.
The empirical values are from the precision data obtained from the
atomic beam laser spectroscopy$^{5)}$. The empirical data exhibit
a conspicuous kink about $^{208}$Pb. The figure also shows the
theoretically obtained isotope shifts for the two forces NL-SH and
NL1. We compare them with those from Skyrme interaction SkM*.
The isotope shifts from NL-SH reproduce the kink very well$^{11)}$.
It is only below A = 198 that the theoretical isotope shifts
show a divergence from the empirical data. This behaviour
is much below the kink and might be attributed to the transitional
behaviour of nuclei in the light Pb isotopes, which is accounted for
in our theory. This region of mass number
usually encounters such effects. The force NL1, on the other hand,
also shows a reasonable kink on the higher side of $^{208}$Pb.
On the lower side, however, NL1 shows a slight divergence from the data
and the slope of the theoretical values is also different from
that of the empirical data. It may be noted that due to inaccurate
description of the other ground-state properties and a very large
asymmetry energy, NL1 is not expected to describe the isotope shifts
adequately. In the same figure we also show for comparison
isotope shifts for SkM* as taken from ref.$^{4)}$. Only the data points
on $^{194}$Pb and $^{214}$Pb are shown, which are representative of the
behaviour of SKM*. On the lighter side, SkM* shows a behaviour similar
to NL-SH. SkM*, however, shows an almost linear function with mass number
and consequently displays a clear divergence from the empirical data on
heavier side.

The kink in the experimental data implies that adding neutrons
to the closed neutron-core of 126 neutrons changes the mean field of
protons which brings about this kink. Attempts have been made to
reproduce this kink using the density-dependent Skyrme forces.
This has been discussed in detail in ref.$^{4)}$. As shown in fig. 2 of
ref.$^{4)}$, all the Skyrme forces e.g. SkM*, Ska and SIII
show a strong deviation from the empirical data in the isotope
shifts for nuclei heavier than $^{208}$Pb. The lighter Pb isotopes could,
however, be described by SkM* and SGII. The binding energies, on the
other hand, show a behaviour opposite to that of isotope shifts.
For example, SkM* reproduces the binding energies of isotopes including and
heavier than $^{208}$Pb. However, it shows a systematic divergence from
empirical binding energies for lighter isotopes.
The disagreement increases on going to the neutron-deficient side.
The other two Skyrme forces show disagreements with the empirical binding
energies on both the sides of the closed shell.
Including all possible corrections beyond the mean field does not
cure the problem. As shown in ref$^{4)}$, Skyrme forces are unable to
describe the binding energies as well as isotopes shifts of nuclei
away from $^{208}$Pb.

The ability to reproduce the kink by the RMF theory and the
failure of the present Skyrme interactions not to be able to do so,
raises some important questions. The shell effects that are inherent
in the structure of nuclei over the periodic table are described
differently by various theories. The kink in the empirical data on
the charge radii of Pb isotopes is one of the aspects which has remained
hitherto unsolved. Presently, the RMF theory succeeds in
accommodating these shell effects which run across the
shell-closure. The shell effects across the magic numbers
play a significant role in astrophysical r-processes.
It should also be noted that  a recent study of nuclei away
from stability line has shown significant differences in the shell
effects predicted by the two approaches$^{12)}$.

The behaviour of the Skyrme mean-field approach and of
the RMF theory towards isotope shifts indicates an important
difference between the two approaches. This difference is
not easy to explain. However, examining the two approaches
we feel that the basic density dependence of the two
interactions and the difference in the saturation mechanism
of the two methods could be at the origin of the difference
in the isotope shifts. Our calculations have shown that the
kink originates from the collective contribution of many
single-particle orbitals to the proton rms radius. The main
contribution comes from the outer orbitals. Another important
aspect that is different is the spin-orbit interaction as mentioned
earlier. Whereas in the RMF
theory, the spin-orbit interaction orginates from the
coupling of $\sigma$ and $\omega$ mesons to Dirac nucleon,
the spin-orbit term in the Skyrme approach is added
phenomenologically. Our study has also shown a slightly different
sequence of the single-particle levels which should be attributed
to the spin-orbit splitting differences in the two approaches.
The spin-orbit splitting is responsible for
putting different orbitals in space and thus determining the
structure of a nucleus. Thus, a difference in the spin-orbit
splitting in the two methods would contribute to the
difference in the isotope shifts of the two approaches.


Very recently it has also been shown that a non-relativistic
reduction of the relativistic Hamiltonian leads to a spin-orbit potential
which exhibits a different isospin-dependence compared with the Skyrme
spin-orbit potential$^{13)}$.
It has turned out, that an isospin-dependent spin-orbit term
in the Skyrme mean field approach can produce
reasonable values for the kink, too. As an example we show in fig. 2
the experimental
values of the isotope shifts in the vicinity of the shell closure
and compare them with SkM* results for two sets of
isospin-admixtures in the spin-orbit term:
\begin{equation}
W_{\tau}(r) = W_{1} \nabla \rho_{\tau} + W_{2} \nabla \rho_{\tau' \ne \tau}\:.
\end{equation}
$W_{1} = 2\, W_{2}$ corresponds to the standard (isospin independent)
spin-orbit
and $W_{1} = 1.05\, W_{2}$ has a strong isospin dependence. All the other
parameters, in particular $W_{2}$, are the same as in SkM* force in both cases.
Further work in this direction is in
progress.\\

\noindent This work is supported by the E.U. HCM programme, contract: EG/ERB
CHBICT-930651
\\

\centerline {\bf References}
\newcounter{ref}
\begin{list}{\arabic{ref})}{\usecounter{ref} \itemsep0cm}
\renewcommand{\baselinestretch}{1.0} \small\normalsize
\frenchspacing
\item B.D. Serot and J.D. Walecka, Adv. Nucl. Phys. {\bf 16} (1986) 1;
       B.D. Serot, Rep. Prog. Phys. {\bf 55} (1992) 1855.
\item P.G. Reinhard, Rep. Prog. Phys. {\bf 52} (1989) 439.
\item D. Vautherin and D.M. Brink, Phys. Rev. {\bf C5} (1972) 626.
\item N. Tajima, P. Bonche, H. Flocard, P.-H. Heenen, and M.S. Weiss,
          Nucl. Phys. {\bf A551} (1993) 434.
\item E.W. Otten, in Nuclear Radii and Moments of Unstable Nuclei, in
          Treaties on Heavy-Ion Science, (ed. D.A. Bromley) Vol. 7
           (Plenum, N.Y. 1988) p. 515.
\item P.G. Reinhard, M. Rufa, J. Maruhn, W. Greiner and J. Friedrichs,
            Z. Phys. {\bf A323} (1986) 13.
\item S.J. Lee, J. Fink, A.B. Balantekin, M.R. Strayer, A.S. Umar,
        P.G. Reinhard, J.A. Maruhn and W. Greiner, Phys. Rev. Lett. {\bf 57}
        (1986) 2916.
\item M.M. Sharma and P. Ring, Phys. Rev. {\bf C45} (1992) 2514.
\item M.M. Sharma, M.A. Nagaragan, and P. Ring, Phys. Lett. {\bf B 312}
      (1993) 377.
\item Y.K. Gambhir, P. Ring, and A. Thimet, Ann. Phys. (N.Y.)
           {\bf 511} (1990) 129.
\item M.M. Sharma, G.A. Lalazissis and P. Ring, Phys. Lett.
           {\bf B317} (1993) 9.
\item M.M. Sharma, G.A. Lalazissis, W. Hillebrandt, P. Ring, Phys. Rev. Lett.
   {\bf 72} (1994) 1431.
\item J. K\"onig et al. in preparation.
\end{list}
\centerline{ Figure Caption}
\begin{itemize}
\item {Fig. 1.}  On the left side (Fig. 1a) the binding energies of Pb isotopes
obtained with the forces NL1 and NL-SH together with the empirical values
(expt.). On the right (Fig. 1b) the isotope shifts with the same forces.
The SkM* values$^{4)}$ and the empirical data are also shown.

\item {Fig. 2.} The isotope shifts of Pb nuclei for two sets of isospin
 admixture in the spin-orbit term of the Skyrme force SkM*.
\end{itemize}
\end{document}